\documentclass[aps,twocolumn,superscriptaddress,showpacs,amsfonts,amsmath,amsthm,amssymb,mathtools,latexsym,amssymb,amsthm,nofootinbib]{revtex4-1}
\usepackage[colorlinks,linkcolor=blue,urlcolor=blue,citecolor=blue,bookmarks,bookmarksnumbered]{hyperref}
\usepackage{bm,braket,dcolumn}
\usepackage{slashed}
\usepackage{graphicx}
\usepackage{epstopdf,epsfig}
\usepackage{enumitem}
\newcommand{\tr}{\text{tr}}
\graphicspath{{pic/}}
\usepackage[usenames,dvipsnames]{color}         
\definecolor{purple}{rgb}{0.5,0,0.5}

\begin{document}
\newcommand*{\PITT}{Department of Physics and Astronomy, University of Pittsburgh, Pittsburgh, Pennsylvania 15260, USA}\affiliation{\PITT}
\title{Entanglement renormalization for chiral topological phases}
\author{Zhi Li}\affiliation{\PITT}
\author{Roger S. K. Mong}\affiliation{\PITT}
\begin{abstract}
We considered the question of applying the multiscale entanglement renormalization ansatz (MERA) to describe chiral topological phases. We defined a functional for each layer in the MERA, which captures the correlation length. With some algebraic geometry tools, we rigorously proved its monotonicity with respect to adjacent layers, and the existence of a lower bound for chiral states, which shows a trade-off between the bond dimension and the correlation length. Using this theorem, we showed the number of orbitals per cell (which roughly corresponds to the bond dimension) should grow with the height. Conversely, if we restrict the bond dimensions to be constant, then there is an upper bound of the height. Specifically, we established a no-go theorem stating that we will not approach a renormalization fixed point in this case.
\end{abstract}
\maketitle

\newcommand{\vx}{\mathbf{x}}
\newcommand{\vy}{\mathbf{y}}
\newcommand{\vz}{\mathbf{z}}
\newcommand{\vi}{\mathbf{i}}
\newcommand{\vj}{\mathbf{j}}
\newcommand{\vk}{\mathbf{k}}
\newcommand{\vl}{\mathbf{l}}
\newcommand{\Gr}{\mathrm{Gr}}
\newcommand\defeq{\mathrel{\stackrel{\makebox[0pt]{\mbox{\normalfont\tiny def}}}{=}}}

\section{Introduction}

Renormalization group (RG) is one of the most important concepts in condensed matter physics for studying long-distance behaviors and topological features.
In real space, RG proceeds by grouping several sites into one effective site, accompanied by a block decimation, a reduction in the local degrees of freedom per site so that it does not increase exponentially with renormalization steps.

Entanglement renormalization \cite{EnRG} provides a concrete realization of a real-space RG for quantum states.
Crucial to entanglement renormalization is the application of \emph{disentanglers} before each coarse-graining step, removing the short-ranged entanglement which then allows the local Hilbert space to decrease.
Entanglement renormalization has been employed in many systems, e.g., to critical phenomena \cite{ex1,ex2,ex3,ex4}, topological ordered phases \cite{meraTO,MERAstringnet}, and quantum fields \cite{Verstraete}.
Applied to a typical (noncritical) state, this RG procedure yields a fixed-point wave function, a state with zero effective correlation length.
These zero-correlation-length states have the property that any connected correlation function is exactly zero beyond some finite distance.
These fixed-point wave functions are often the ``model wave functions'' for the corresponding topological phase \cite{SchuchCiracPerezGarciaPEPS,LevinWen,LUT}.

The multiscale entanglement renormalization ansatz (MERA) \cite{VidalMERA} is a tensor network description of the entanglment RG procedure.
By keeping track of the disentanglers and decimations at each RG step, a MERA network can be reversed to recover the original quantum state from a coarse-grained one.
In other words, a MERA, considered as a quantum circuit, can be used to recover the short-distance physics from the long-distance physics.

In this Rapid Communication, we investigate the possibility to use MERA to describe chiral topological states.
We show that there are no \emph{IR fixed points} to chiral Chern insulators on lattices;
any Chern insulator on a lattice with local dimension $D$ must admit a finite correlation length $\xi$, and we argue that there must be a trade-off between $D$ and $\xi$.

Specifically, we consider a fermionic Gaussian MERA along with an IR wave function for a Chern insulator.
We define a functional $L$ for each layer in the MERA which captures its correlation length, and rigorously prove that it obeys monotonicity with respect to adjacent layers.
In addition, we prove the existence of a lower bound for $L$ when the Chern number is nonzero, and such a bound is a decreasing function of the bond dimension.

Our results can be interpreted as follows:
Consider a wave function for the ground state of a Chern insulator $\psi_0$, undergoing a series of entanglement renormalization steps to generate coarse-grained wave functions $\psi_1, \dots, \psi_n, \dots$.
Naturally, as in any RG procedure, we expect the correlation length $\xi(\psi_n)$ to decrease exponentially with the number of renormalization steps $n$.
Our results imply that either we need more orbitals per cell as we continue the renormalization process, or the Chern number must change for some large $n$.
The former case implies that the bond dimension of a MERA must grow with the number of layers, while the latter scenario implies that the RG procedure has failed to capture the topological properties of the state.

This Rapid Communication is organized as follows. In Sec.~\ref{sec-mera}, we give a short review of MERA and define the notation used here. In Sec.~\ref{sec-theorem}, we state the main theorem and discuss its physical implications. Then, in Sec.~\ref{sec-sketch} (and Supplemental Material \cite{supp} for details), we prove this theorem. Finally, in Sec.~\ref{sec-discuss}, we give some discussions and outlooks.

\section{Entanglement renormalization and MERA}\label{sec-mera}
In this section we briefly describe the entanglement renormalization and multiscale entanglement renormalization ansatz (MERA).
While our results apply to MERA in general dimensions, here we review one-dimensional (1D) MERA for simplicity. 

\begin{figure}[h]
	\centering
	\includegraphics[width=1.0\columnwidth]{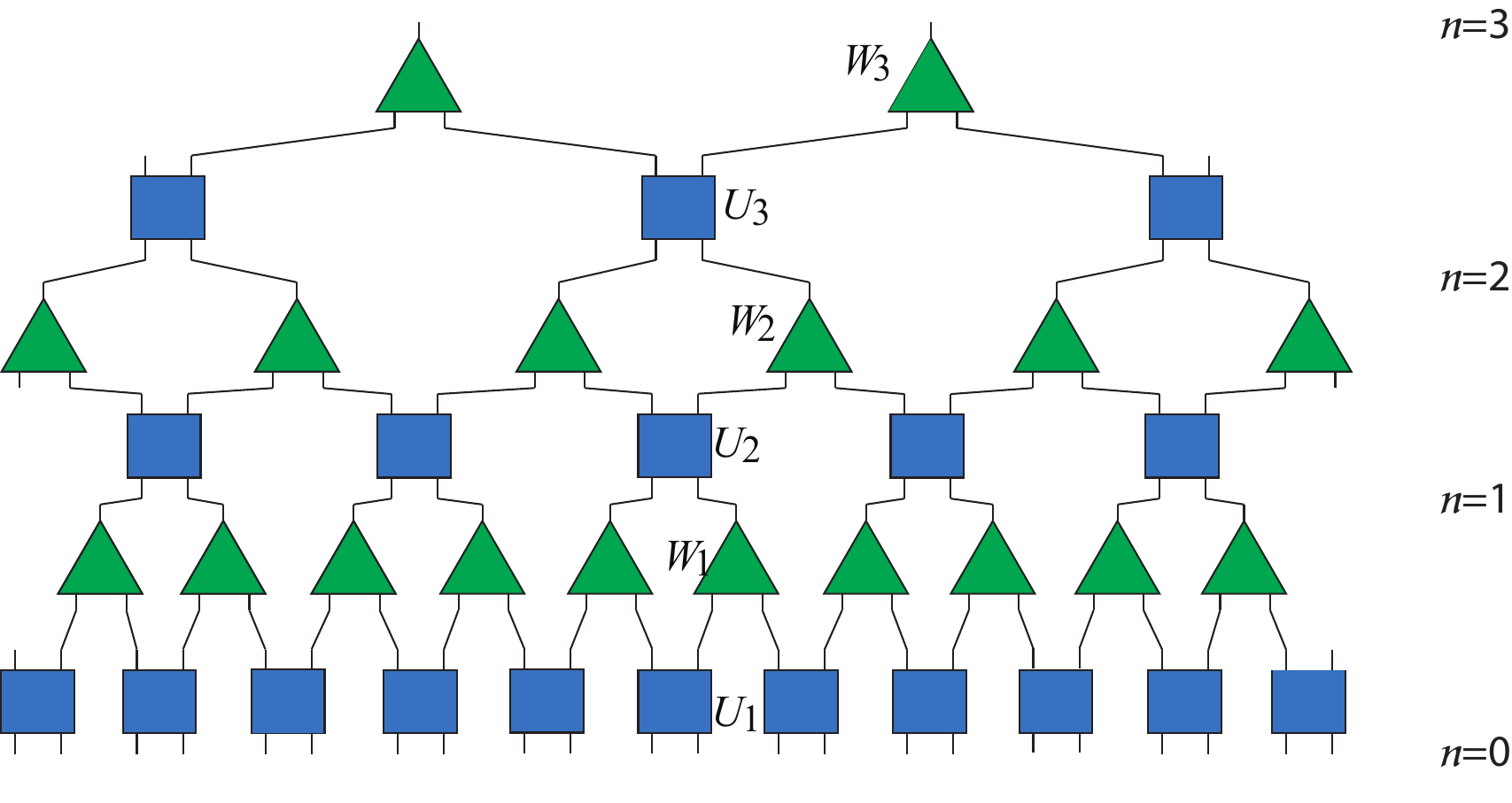}
	\caption{
		A three-layer binary MERA in 1D.
		Putting an IR (coarse-grained) wave function at the top generates a UV wave function at the bottom.
		The blue blocks are the disentanglers ($U_1,U_2,U_3$), and the green triangles are the isometries ($W_1,W_2,W_3$).
		Each of the elements can be thought of as operators; disentanglers are two-site unitary gates, and isometries create additional lattice sites.
	}\label{pic-mera}
\end{figure}

We view entanglement renormalization as a process which takes a short-distance (UV) description of a system to a long-distance (IR) description.
For this work, we want to restrict to entanglement renormalization processes that are \emph{reversible}, in the sense that the UV limit can be recovered exactly from the IR. In other words, the MERA is an exact representation of the UV wave function, a quantum circuit which allows the UV physics (e.g., correlation functions) to be reconstructed from the IR physics (i.e., symmetry breaking, topological order).


In the ordinary real-space renormalization, we simply group several sites into one effective site, resulting in a tree tensor network (TTN), as shown in Fig.~\ref{pic-mera} if we ignore the blue rectangular blocks. Here, the coarse-graining process is represented by the green triangles, called \emph{isometries}, denoted by $W$.
Each line represents a physical degree of freedom (i.e., a spin on a lattice site). The layers (labeled by $n$, counted from below, as shown in the figure) represent intermediate steps of the RG process. Regarded as a quantum circuit (topdown), each green triangle enlarges the Hilbert space, and describes an isometric embedding from layer $n+1$ to layer $n$.

In general, one needs more and more ``local degrees of freedom" (i.e., the local Hilbert space grows with each iteration) to compensate the coarse graining due to the entanglement structure (see Ref.~\onlinecite{MERA-intro} for an argument using the entanglement entropy).
The way to fix this problem is to apply ``disentanglers" between coarse-graining steps to reduce the cross-site entanglement.
They are simply some unitary transformations among adjacent sites, denoted by $U$, represented by the blue rectangular blocks in Fig.~\ref{pic-mera}.
The resulting tensor network is called the multiscale entanglement renormalization ansatz (MERA).

To maintain translational invariance, we will assume the disentanglers and isometries within one layer are the same (but they may differ from layer to layer). Then the states in all layers are translationally invariant if and only if the state in at least one layer is translationally invariant (with different periods in general). Formally speaking, a (translationally invariant) MERA with $L$ layers is specified by the following data:
\begin{itemize}[topsep=1mm]
\setlength\itemsep{1mm}
\setlength\parskip{0mm}
\item	isometries $W_n$,
\item	disentanglers $U_n$,
\item	bond dimensions $D_n$,
\item	top-level wave function $\psi_L$.
\end{itemize}
Note that the bond dimension $D$ referred to here is the noninteracting one, which is equal to the number of orbitals in the site.
The conventional (interacting) bond dimension for a tensor network is the dimension of the local Hilbert space, which is equal to $2^D$ if the physical degree freedom in a site is a qubit.

The generalization to higher dimensions is evident \cite{MERA2D0,MERA2D}. Note that even in 1D, we can have different types of MERA: For example, we may construct a ternary MERA where each isometry has three legs \cite{ex3}. In two dimensions (2D) or more, the choices of isometries and disentanglers are more diverse.

\section{Main theorem: statement and implications}\label{sec-theorem}

We would like to see what will happen if we want to apply MERA to describe chiral states. Here, we focus on Chern insulators living on 2D lattices, and the generalization to higher dimensions is straightforward.

We will call the minimal geometrical translationally invariant unit as a site. The sites must form a $\mathbb{Z}^2$ lattice.
There may be additional degrees of freedom per site (such as sublattice structure, orbitals, spins), which we collectively refer to as orbitals.
The total number of orbitals per site is what we call the bond dimension $D$, so there is a vector of annihilation operators for each site $\vx$:
$\phi_\vx=(\phi_{\vx1},\dots,\phi_{\vx D})$.

Here, we only consider translationally invariant states. We will call the minimal translationally invariant unit for a state as a cell, denoted by $C$.
In general, a cell may contain multiple sites,
\begin{align*}
	\textrm{orbitals} \subseteq \textrm{sites} \subseteq \textrm{cell} .
\end{align*}

As usual, one can define the correlation matrix $P_{\vx,\vy} = \braket{\phi^\dag_\vx \phi_\vy}$ for each layer, where $\vx$ and $\vy$ label sites.
For a noninteracting fermionic system, the $P$ matrix is a projector onto filled bands (see Ref.~\cite{supp}), and encodes all the information of the state, including its topological properties.

We define a functional $L$ for each layer as
\begin{equation}\label{eq-defB}
L(P)=\frac{1}{|C|}\sum_{\vy\in \text{C}}\sum_{\vx\in\mathbb{Z}^2} a_{\vx-\vy}||P_{\vx,\vy}||^2.
\end{equation}
Here, $|C|$ is the size of the unit cell (the number of sites in $C$), $\{a_\vx\}$ are nonnegative constants to be specified below, and $||\cdot||$ is the Hilbert-Schmidt norm, defined as
\begin{equation}
	||A||_{HS}^2=\tr(A^\dagger A).
\end{equation}
For gapped states, $P_{\vx,\vy}$ decays at least exponentially \cite{Hastings_decay} with respect to $|\vx-\vy|$, so we demand $a_\vx$ to be asymptotically polynomial to guarantee the convergence.
The factor $1/|C|$ makes $L$ independent of the choice of the unit cell.
It is appropriate to think of $L$ as a proxy for the correlation length (see Sec.~\ref{sec-discuss} for details).

\textbf{Theorem.} For each number $s>2$, there exists a constant $A>1$ and a function $a_\vx$ such that $a_\vx \rightarrow |\vx|^{s}$ asymptotically and that the functional $L$ satisfies the following properties:
\begin{enumerate}
\item
(Monotonicity) $\forall n$, we have $L^{(n)}\geq A L^{(n+1)}$. Here, $L^{(n)}$ represents the value of $L$ for $n$th layer.

\item
(lower bound) If the Chern number $c\neq 0$, then $L$ has a strictly positive lower bound $\epsilon$.  The bound $\epsilon$ will depends on the Chern number $c$ and the number of orbitals per cell $N=|C|D$. Note that although $L$ does not dependent on how we identify the unit cell, $N$ does. The strongest lower bound is given by the minimal unit cell. 
\end{enumerate}

The choice of $a_\vx$ is as follows: We pick a finite region $F_{s,A}\subset\mathbb{Z}^2$ (specified in Ref.~\cite{supp} where we show the existence of such a region), which includes the origin, then define
\begin{equation}\label{eq-defax}
	a_\vx=\begin{cases}
	0,&\vx\in F_{s,A}\\
	|\vx|^{s},&\vx\notin F_{s,A}
	\end{cases}.
\end{equation}

Before proving the theorem, we discuss its physical interpretations and implications. 

First of all, the existence of a lower bound of $L$ shows that topology imposes a restriction on the ``correlation length." It obvious that $\epsilon$ is a \emph{decreasing} function of $N$ (because by definition it is a lower bound and we can embed a small cell into a larger one by adding empty bands). This means there is a trade-off between the bond dimension and the ``correlation length."
 
Now, we assume there is a MERA (finite layer or infinite layer) generating a given chiral state.
From monotonicity, $L^{(n)}\leq A^{-1}L^{(n-1)}\leq\cdots\leq A^{-n}L^{(0)}$ for all $n$.
The Chern number, denoted by $c$, must be the same for each layer \cite{Shinsei}.
So we have
\begin{equation}\label{eq-Dbound1}
A^{-n}L^{(0)}\geq\epsilon(c,N_n).
\end{equation}
Since $\epsilon$ is a decreasing function of $N$, the above inequality gives us a lower bound of $N_n$,
\begin{equation}\label{eq-Dbound2}
N_n\geq\epsilon^{-1}(c,A^{-n}L^{(0)}),
\end{equation}
where $\epsilon^{-1}(c,\cdot)$ is the inverse function of $\epsilon(c,N)$ with respect to the second argument; this lower bound is an \emph{increasing} function of $n$. 

Physically, it means that for a given chiral state $\psi_0$ (with $c\neq0$) at the bottom, there will be a lower bound of orbitals per cell $N$ for each layer, and the bound will increase with the MERA's depth $n$. (Note that this statement is only about the lower bound of $N_n$; for a specific MERA, the actual number $N_n$ in each layer does not necessarily increase with the layer index.) Equivalently, given a chiral state $\psi_0$, if we want to use an MERA with $(n+1)$ layers to generate it, we need in general more orbitals per cell on the top layer compared with an $n$-layer MERA. In particular, if we want the bond dimension to be asymptotically constant, Eq.~\eqref{eq-Dbound1} gives us an upper bound of the depth $n$. So we obtain the following:

\textbf{No-go theorem.} No infinite-layer MERA with asymptotically constant bond dimension could represent a gapped translationally invariant chiral state.

On the other hand, let us fix the bond dimension on the top layer; then Eq.~\eqref{eq-Dbound1} implies that the value of $L^{(0)}$ for the UV layer will diverge with the number of layers $n$ and hence the wave function $\psi_0$ must also have a diverging (with respect to $n$) correlation length.

In the case of infinite-layer MERA with an asymptotically constant bond dimension, the same logic show that not only is it impossible to represent a chiral state (the above no-go theorem), no such infinite-layer MERA can even provide a good approximation in the sense of $L$.
Note that it might be possible to approximate a chiral state in other senses \cite{chiralMERA}, however, the situation is similar to the projected entangled pair state (PEPS) case \cite{cirac-peps}: Free fermionic PEPS cannot correspond to the exact ground states of gapped, local parent Hamiltonians, but they can nevertheless provide an approximation. The difference between a chiral PEPS and the exact state is the ``tail behavior" (for example, power versus exponential), which is hard to distinguish by a naive norm, but can be distinguished by our $L$.



\section{Sketch of the proof} \label{sec-sketch}
Now we sketch the proof of this theorem. The details will be given in the Supplemental Material \cite{supp}. 

The proof of monotonicity (Theorem~1) is straightforward. To keep the basic idea as clear as possible, we will use words such as ``exists a constant" and ``when $|\vx|$ is large enough." We use the standard 2D MERA for an example, since the general case is similar. 

Since the second-quantization operator $\phi_\vx^{(n)}$ in the $n$th layer is linearly related to those in the $(n+1)$th layer by $W$ and $U$, we can represent the correlation matrix $P^{(n+1)}$ using $P^{(n)}$.
The tensors in a MERA are local: Each block ($W$ or $U$) has at most four legs in each side, so it is easy to show $\vx$ in the $(n+1)$th layer only talks to $2\vx+\vi$ in the $n$th layer so that $P^{(n+1)}_{\vx,\vy}$ is only related to $P^{(n)}_{2\vx+\vi,2\vy+\vj}$, where $\vi$ and $\vj$ are valued in a finite set.
Plugging the linear relation between $P^{(n+1)}$ and $P^{(n)}$ into Eq. (\ref{eq-defB}), one obtains an inequality with the following form,
\begin{equation}\label{eq-sk1}
L^{(n+1)}\leq C_1 \sum a_{\vx-\vy}||P^{(n)}_{2\vx+\vi,2\vy+\vj}||^2,
\end{equation}
for some constant $C_1$. 

To prove $L^{(n)}>AL^{(n+1)}$ for some $A$, we need 
\begin{equation}\label{eq-sk2}
	a_{\vx-\vy}< C_2 a_{(2\vx+\vi)-(2\vy+\vj)},
\end{equation}
so the right-hand side of Eq.~\eqref{eq-sk1} goes to $L^{(n)}$. 
This is obvious from Eq.~\eqref{eq-defax} provided that $s$ is large enough.

In order to prove the existence of the lower bound (Theorem~2), we proceed in two steps. 

First, we prove that $L\neq 0$ as long as the state is chiral ($c\neq 0$) no matter how we choose $a_\vx$. Recall the definition of $L$ and $a_\vx$ in Eqs.~(\ref{eq-defB}) and (\ref{eq-defax}); what we need is for any finite region $F_{s,A}$, $P_{\vx,\vy}$ cannot simultaneously vanish for all $\vx-\vy\notin F_{s,A}$. This is where algebraic geometry tools are used. Roughly speaking, a counterexample will induce an ``algebraic bundle" over the torus, which must be trivial (see Proposition 1 in the Supplemental Material \cite{supp}). Physically, this means although the correlation is short ranged in the sense that it decays exponentially, it cannot be strictly local (as in many zero-correlation-length ``model wave functions'').

Second, we use a continuity argument to show the infimum (best lower bound) of $L(P)$ must also be positive. If not, there will be a sequence of maps such that $L(P)\to 0$.
A limit $\tilde{P}$ of a subsequence will satisfy $L(\tilde{P})=0$ (for a slightly larger $F_{s,A}$), hence the Chern number $c(\tilde{P})=0$ according to the first step. However, the Chern number, as an integer, should not jump when taking the limit, which provides a contradiction.

\section{Discussion}\label{sec-discuss}

While our results are phrased in terms of a MERA tensor network, the statements we make are applicable to entanglement renormalization as a whole.
Particularly, \emph{entanglement renormalization fails for a Chern insulator on a lattice}, provided one demand the RG procedure is reversible.

Our proof of Theorem~2 is based on some algebraic geometry tools. It will be interesting to see if similar tools can be used to solve other problems. On the other hand, the proof is not constructive: It does not provide an explicit expression for the lower bound.
However, one can give a very rough estimation of $L(P)$ and the lower bound function $\epsilon(c,N)$ as follows. 

Consider the case where $|C|=1$, $N=D$ (the general case will be similar). 
Let us group $l^2$ lattices into an effective cell, so that $N\to l^2 N$. One gets a new series $P'_\vx=(P_{l\vx+\vi-\vj})$, where $\vi$, $\vj$ are labels in the new cell $C'$ (now with linear size $l$).
For an $\vx$ at the boundary of the region $F$, due to the fast decay of $P_\vx$, we can apply the saddle-point method to estimate $||P'_\vx||^2$,
\begin{equation}
||P'_\vx||^2=\sum_{\vi,\vj}||P_{l\vx+\vi-\vj}||^2\sim ||P_{l\vx+\vi_0-\vj_0}||^2\sim e^{-\alpha|F|l}.
\end{equation}
Here, the first $\sim$ is because only the largest element (when $\vi,\vj$ are at some corners of the new cell) in the summation contributes, the second $\sim$ assumes $P_\vx$ indeed decays exponentially with $\alpha$ as the decay rate. $|F|$ is the radius of $F$. Also by the saddle-point approximation, one has:
\begin{equation}\label{eq-estimate1}
L(P')\sim \sum_{\vx\in\partial(F)}|\vx|^{s}||P'_\vx||^2 \sim |\partial F||F|^{s}e^{-\alpha|F|l}.
\end{equation} 
Here, $|\partial F|$ is the perimeter of the boundary. 
Equation~\eqref{eq-estimate1} is valid when the linear size of $P$ is $l^2N$, so in general for $P$ of linear size $N$, we have $L(P)\sim e^{-\beta\sqrt{N}}$, where $\beta$ is another constant. In particular, $\epsilon(c,N)\lesssim e^{-\beta\sqrt{N}}$. 

This is just the crudest estimation. One could obtain a better estimation given a faster (than exponential) decayed. From another point of view, this argument gives a refinement of Proposition 1 in Sec.~\cite{supp}: Not only $P_\vx$ cannot simultaneously vanish for large $\vx$, but it cannot decay faster than a bound set by $\epsilon(c,N)$. We do not know what $\epsilon(c,N)$ is exactly. If we assume $\epsilon(c,N)\sim e^{-\beta\sqrt{N}}$, then Eq.~\eqref{eq-Dbound2} tells us $N_n\gtrsim n^2$ (ignore all the coefficients). Note again that this is not a proven bound of $N_n$: If $\epsilon$ decays faster,  $N_n$ grows more slowly.

Part of our conclusions can be understood from another way.
It was shown in Ref.~\onlinecite{mera-peps} that a MERA with a bounded bond dimension $D_M$ (they use $\chi_M$) can be mapped into a PEPS with a bounded bond dimension $D_P$ which is a polynomial of $D_M$ and independent of the system size and the number of layers (they call this property \emph{efficiency}). One can generalize their proof to the case of infinite-size MERA, and hence obtain an infinite-size PEPS with bounded bond dimension and no ``input" on the top. However, according to Refs.~\cite{dubail-read,cirac-peps}, PEPS (with no input) cannot generate exact ground states of gapped, local parent Hamiltonians.
So, we conclude that no infinite-layer MERA with a bounded bond dimension could represent a gapped chiral state. Compared to the above argument, our treatment here emphasizes the renormalization point of view from where MERA originates. 

At last, we mention some possible generalizations. Here, we focused on the 2D noninteracting translationally invariant chiral states. In $d$ dimension, we need $s>d$ to guarantee both the monotonicity and a convergence in the proof of the theorem. 

One possible generalization is to the case without translational invariance. Here, the state is also determined by the correlation matrix $P$, but one cannot use a Fourier transformation and band structures due to the lack of translational invariance. Instead, one should, for example, proceed in the spirit of Refs.~\cite{Kitaev-chern,Hastings-chern} to define the Chern number. The first part of our theorem is still valid with almost no changes in the proof.
It is plausible that a  construction similar to our functional $L(P)$ also has a nonzero lower bound and one can proceed similarly to show the obstruction provided by the topology. 

The generalization to the interacting case is certainly worth exploring. 
We conjecture that the same result holds in the presence of any chiral anomaly.
In particular, the $\mathrm{U}(1)$ chiral anomaly [e.g., in the case of the $\mathrm{U}(1)$ boson symmetry-protected topological (SPT) phase \cite{Chen}, which manifests itself in the form of a quantized Hall conductance] would prevent a lattice fixed-point IR state to be constructed.
In addition, the gravitational chiral anomaly, which arises from a nonzero chiral central charge, may also provide such an obstruction.

\begin{acknowledgements}
We are grateful to Spiros Michalakis and Michael Zaletel for discussions. We thank the anonymous referees for suggestions and discussions on this paper.
\end{acknowledgements}

\bibliography{merabound}

\end{document}


\newcommand*{\PITT}{Department of Physics and Astronomy, University of Pittsburgh, Pittsburgh, Pennsylvania 15260, USA}\affiliation{\PITT}
\title{Supplemental Material for ``Entanglement renormalization for chiral topological phases"}
\author{Zhi Li}\affiliation{\PITT}
\author{Roger S. K. Mong}\affiliation{\PITT}
\begin{abstract}
This note is a supplemental material for ``Entanglement renormalization for chiral topological phases", where we present the proof of our theorems. In this first part, we present the explicit construction of our functional and address its monotonicity. In the second part, we prove the existence of the lower bound. Essential to the prove are some algebraic geometry tools.
\end{abstract}
\maketitle

\newcommand{\vx}{\mathbf{x}}
\newcommand{\vy}{\mathbf{y}}
\newcommand{\vz}{\mathbf{z}}
\newcommand{\vi}{\mathbf{i}}
\newcommand{\vj}{\mathbf{j}}
\newcommand{\vk}{\mathbf{k}}
\newcommand{\vl}{\mathbf{l}}
\newcommand{\Gr}{\mathrm{Gr}}
\newcommand\defeq{\mathrel{\stackrel{\makebox[0pt]{\mbox{\normalfont\tiny def}}}{=}}}

\section{Definition of functional $L(P)$ and monotonicity}

\newcommand{\R}{\alpha}
\renewcommand{\S}{\beta}
\newcommand{\T}{\gamma}
\newcommand{\s}{\beta}
\renewcommand{\t}{\gamma}

\begin{figure}[h]
  \centering
  \includegraphics[width=0.5\textwidth]{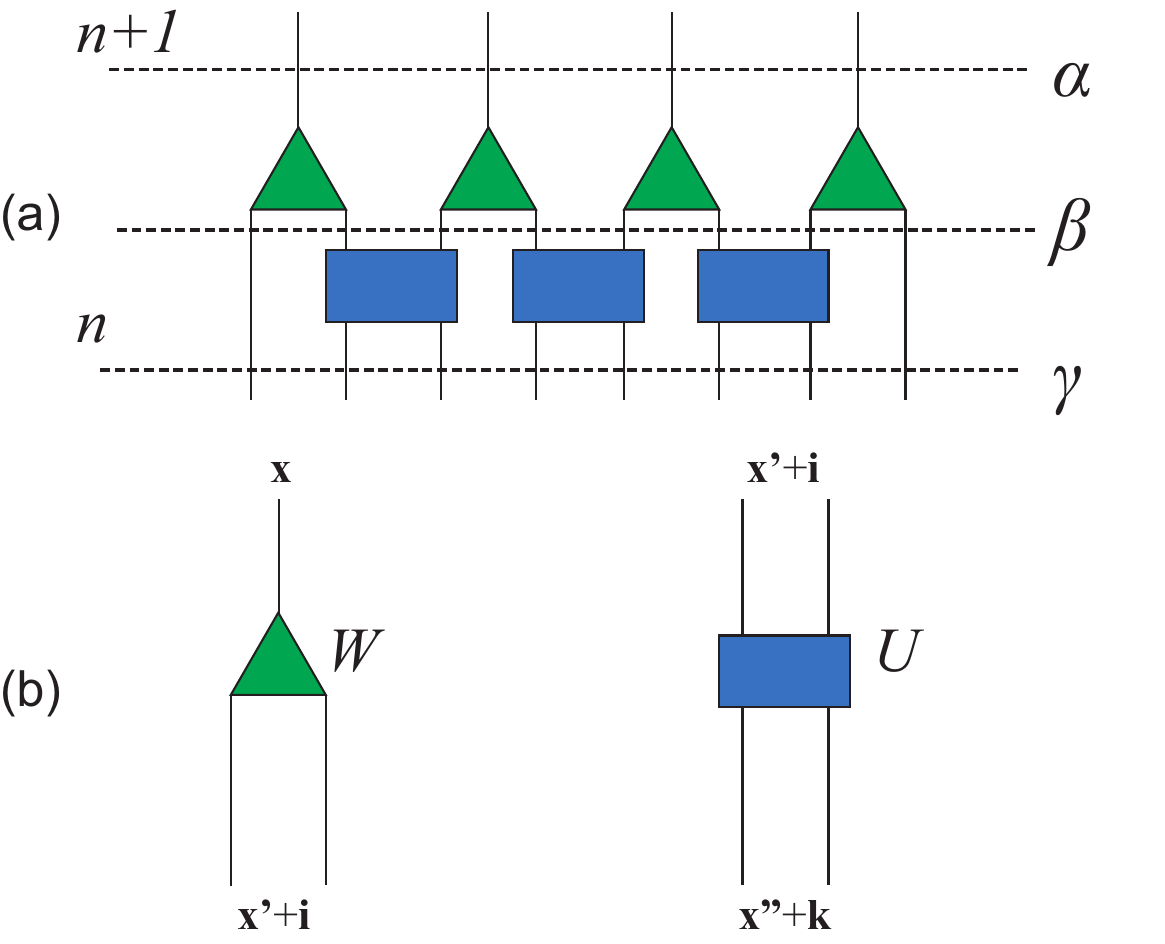}\\
  \caption{The illustration of our notation in 1D case. (a) We denote those three relavant layers to be $\R,\S,\T$. (b) The definition of $\vx'+\vi$. (c) The definition of $\vx'+\vi$ and $\vx''+\vk$. In this case, $\vx,\vy$ are valued in $\mathbb{Z}$, $\vx'=2\vx$, $\vi,\vj,\vk,\vl$ are valued in \{0,1\}}\label{pic-mono}
\end{figure}

In order to proof the monotonicity of $L(P)$, we compare $P_{\vx,\vy}^{(n+1)}$ with $P_{\vx,\vy}^{(n)}$. We denote the $(n+1)$th layer to be $\alpha$, $n$th layer to be $\gamma$, the layer between them (below the isometry, above the disentangler) to be $\beta$. See Fig.~\ref{pic-mono} for illustration.

For the isometry $W=(W_\vi)$ between $\R$ and $\S$, we have
\[\phi_\vx^\R=\phi_{\vx'+\vi}^\S W_\vi,\]
so $P^\R_{\vx,\vy}=\langle\phi^{\R\dagger}_{\vx}\phi^{\R}_{\vy}\rangle=W^\dagger_\vi P^{\S}_{\vx'+\vi,\s(\vy)+\vj} W_\vj=W^\dagger P^{\S}_{\vx,\vy} W$. Here $\vx'$ is a representative point under the isometry starting with $\vx$ (for the standard binary MERA in 2D, we can just choose $\vx'=2\vx$). $W$ is an isometry in the sense that $W^\dagger W=1_\R$ and $WW^\dagger$ is a projection in the local Hilbert space correspond to $\vx'+\vi$. $P^\S_{\vx,\vy}$ is the matrix with elements $P^{\S}_{\vx'+\vi,\s(\vy)+\vj}$.

By the definition of the Hilbert-Schmidt norm, we have
\begin{equation}\label{eq-rs}
\begin{aligned}
||P^\R_{\vx,\vy}||^2&=||W^\dagger P^{\S}_{\vx,\vy} W||^2=\text{tr}(P^\S_{\vx,\vy}WW^\dagger P^{\S\dagger}_{\vx,\vy}WW^\dagger)\\
&\leq ||P^\S_{\vx,\vy}WW^\dagger||\cdot||WW^\dagger P^\S_{\vx,\vy}||\leq ||P^\S_{\vx,\vy}||^2\\
&=\sum_{\vi\vj}||P^\S_{\vx'+\vi,\s(\vy)+\vj}||^2.
\end{aligned}
\end{equation}
Here, we're using the Cauchy inequality for Hilbert-Schmidt norm and the fact that $WW^\dagger$ is a projection (hence $||WW^\dagger P||\leq ||P||$). 

In practice, in a MERA we have
\[\phi^\S=(\phi^\R,\ast) \tilde{W}^\dagger=\tilde{\phi}^\R\tilde{W}^\dagger,\]
where $\tilde{W}$ is a unitary augmentation of $W$, where the ``$\ast$" denotes some local ancillas with no correlation with $\phi^R$ and other stars, so $\tilde{W}P^\S_{\vx,\vy}\tilde{W}^\dagger=\tilde{P}_{\vx,\vy}^\R$. Hence Eq. (\eqref{eq-rs}) is actually an equality when $\vx\neq\vy$. However, we don't need this result.

For the disentangler $U$ between $\S$ and $\T$, we have, similarly,
\begin{equation}\label{eq-RS1}
\sum_{\vi\vj}||P^\S_{\vx'+\vi,\vy'+\vj}||^2= \sum_{\vk\vl}||P^\T_{\vx''+\vk,\vy''+\vl}||^2
\end{equation}
where $\vx''$ is a representative point under the disentangler that contains $\vx'$. Here, the summation of $\vi,\vj$ and $\vk, \vj$ is over the leg of the two disentanglers, see Fig.~\ref{pic-mono}(c). 

Now we can compare $L^\R$ and $L^\T$ as follows:
\begin{equation} \label{eq-RS2}
\begin{aligned}
L^\R&=\frac{1}{|C_{n+1}|}\sum_{\vy\in C_{n+1}}\sum_{\vx\in \mathbb{Z}^2}a_{\vx-\vy}||P^\R_{\vx,\vy}||^2\\
&\leq\frac{1}{|C_{n+1}|}\sum_{\vy\in C_{n+1}}\sum_{\vx\in \mathbb{Z}^2}a_{\vx-\vy}\sum_{\vi\vj}||P^\S_{s(\vx)+\vi,s(\vy)+\vj}||^2\\
&=\frac{w}{|C_{n}|}\sum_{\vy ' \in C_n}\sum_{\vx ' \in \mathbb{Z}^2}a_{\vx-\vy}||P^\S_{\vx ',\vy '}||^2.
\end{aligned}
\end{equation}
Here $\vx ', \vy '$ denote bonds in the $\S$ layer, $w$ is the number of components under the isometry $W$ (for standard binary MERA in 2D, $w=2^2=4$), so that $|C_n|=w|C_{n+1}|$. Note that we have the freedom to enlarge the unit cell so we can assume this relation without loss of generality. We can also assume that the cell $C_n$ exactly contains several disentanglers.

Denote $T_{\vx'}$ to be the disentangler that $\vx'$ belongs to. We would like to have the following property for each pair of disentanglers ($T_{\vx'},T_{\vy'}$):
\begin{equation} \label{eq-if1}
a_{\vx-\vy}\leq \frac{1}{M} a_{\vx''-\vy''} \text{~for~} \forall \vx,\vx''/\vy,\vy'' \text{~connected to~} T_{\vx'}/T_{\vy'}.
\end{equation}
($\vx$ is connected to $T_{\vx'}$ means the isometry under $\vx$ has common legs with $T_{\vx'}$; $\vx''$ is connected to $T_{\vx'}$ means $\vx''$ is a leg of $T_{\vx'}$.) If so, from Eq. (\ref{eq-RS1}) we know
\begin{equation}
\sum_{\vi,\vj}a_{\vx-\vy}||P^\S_{\vx',\vy'}||^2\leq\frac{1}{M}\sum_{\vk,\vl}a_{\vx''-\vy''}||P^\T_{\vx'',\vy''}||^2\\
\end{equation}
so we can continue as follows:
\begin{equation}
\begin{aligned}
L^\R&\leq \frac{w}{M|C_n|}\sum_{\vy'' \in C_n}\sum_{\vx'' \in \mathbb{Z}^2}a_{\vx''-\vy''}||P^\T_{\vx'',\vy''}||^2=\frac{w}{M} L^\T.\\
\end{aligned}
\end{equation}
As long as 
\begin{equation}\label{eq-if2}
A\defeq\frac{M}{w}>1,
\end{equation}
we will get the desired inequality.

So the question is to choose $\{a_\vx\}$ such that Eq. (\ref{eq-if1},\ref{eq-if2}) hold. This is always possible. For example, for the standard MERA, $\vx'=2\vx$, $w=4$. One can easily see that the distance between $\vx''$ and $2\vx$ at most by $2\sqrt{2}$ so that the distance between $(\vx''-\vy'')$ and $2(\vx-\vy)$ is at most $4\sqrt{2}$. We demand $a_\vx=|\vx|^{s}$ asymptotically so that 
\begin{equation}
\lim_{|\vx-\vy|\to\infty} \frac{a_{\vx''-\vy''}}{a_{\vx-\vy}}=2^s.
\end{equation}
We demand $s>2$, hence for $\forall M$ such that $w=4<M<2^s$, Eq. (\ref{eq-if1}) hold when $|\vx-\vy|$ is large (depends on $M$) enough. Problems may happen when $|\vx-\vy|$ is small, but we can simply demand $a_\vx=0$ for $\vx$ in some finite region $F$ to fix the problem. 

The region $F$ in general depend on $s$ and $A$. Indeed, since $|(\vx''-\vy'')|\geq |2(\vx-\vy)|- 4\sqrt{2}$, Eq. (\ref{eq-if1}) holds when $|\vx-\vy|\geq\frac{4\sqrt{2}}{2-(4A)^{\frac{1}{s}}}$. So we can simply demand $F$ contains the disc $D(0,\frac{4\sqrt{2}}{2-(4A)^{\frac{1}{s}}})$

\section{Existence of the lower bound} \label{sec-lowerbound}
First, we recall how the Chern number (or other topological invariants) is defined from the correlation matrix $(P_{\vx,\vy})$. 

We start with a special case: when we have a translationally invariant Hamiltonian,
\[\mathcal{H}=\sum_{\vx,\vy}\phi_\vx^{T\dagger} H_{\vx-\vy}\phi_\vy^T= \int\frac{d^2\vk}{(2\pi)^2}\phi_\vk^{T\dagger} H(\vk)\phi^T_\vk,\]
and the state is the ground state. Here the integral is taken over the Brillouin zone $T^2$, each $H(\vk)$ is a $D\times D$ Hermitian matrix with $q$ negative eigenvalues, corresponds to the $q$ filled bands. The assignment ``$\vk\to\text{filled subspace of } H(\vk)$" (equivalently, a map from $T^2$ to the Grassmanian manifold $\Gr(D,q)$) gives us a vector bundle over $T^2$. The Chern number is a characteristic number \cite{MilnorBook} of this bundle, which is used to classify topologically inequivalent bundles.

It is straightforward to show that the ground state correlation function is roughly equal to the projection to the filled bands:
\begin{equation}
P_{\vk\vk'}=\langle\phi_\vk^\dagger\phi_{\vk'}\rangle = (2\pi)^2\delta(\vk-\vk')\mathcal{P}_-(H(\vk)),
\end{equation}
where $\mathcal{P}_-(H)=\oplus_{\lambda<0}P_\lambda$, $P_{\lambda}$ is the projection matrix on the eigenspace of $H$ with eigenvalue $\lambda$. 


In general cases where the (minimal) unit cell $C$ may contain several sites and the Hamiltonian is not given in advance, we can proceed exactly as above. Regard the matrix $(P_{\vx,\vy})$ as a block matrix $(P_{\bar{\vx}\lambda,\bar{\vy}\mu})$ (here $\lambda,\mu$ are labels in a unit cell, $\lambda,\mu\in\{1,2,\cdots,|C|\}$), which only depends on $\bar{\vx}-\bar{\vy}$ and $\lambda,\mu$. Then taking the Fourier transform with respect to $\bar{\vx}-\bar{\vy}$, one obtains projection matrices $P(\vk)=(P_{\lambda,\mu}(\vk))$ and hence a map 
\[P:T^2\to\Gr(N,q),\]
where $N=|C|D$. The Chern number $c$ of the state is just the Chern number of this map or the Chern number of the corresponding vector bundle $P^*(\tau)$, the pullback \cite{MilnorBook} of the tautological bundle over $\Gr(N,q)$.

As a note to be used later, here we have embeded $\Gr(N,q)$ into $M(N,\mathbb{C})$, the space of $N\times N$ complex matrices. Indeed, a point of $\Gr(N,q)$ corresponds to a $q$-dimensional subspace, which uniquely corresponds to the orthogonal projection matrix onto this subspace.

Go back to our theorem. As a first step, we prove the following weaker statement which ensures $L(P)\neq 0$ for chiral state.

\textbf{Proposition 1.} If $P_{\vx,\vy}=0$ for $\forall \vx,\vy$ such that $|\vx-\vy|$ is large enough, then $c(P)=0$.

\textbf{Proof:} We want to show the existence of $q$ everythere-linear-independent global sections of $P^*(\tau)$, hence the bundle $P^*(\tau)$ is trivial. 

To proceed, write the matrix-valued map $P(\vk): T^2\to \Gr(N,q)\subseteq M(N,\mathbb{C})$ in components $p_{ij}(\vk)$ ($i,j\in\{1,2,\cdots,N\}$). Denote $x=e^{ik_1}$, $y=e^{ik_2}$, where $\vk=(k_1,k_2)$. The condition ``$P_{\vx,\vy}=0$ for sufficiently large $|\vx-\vy|$" is now equivalent to ``each $p_{ij}(x,y)$ is a Laurent polynomial of $x,y$", i.e: $p_{ij}\in R\defeq\mathbb{C}[x,x^{-1},y,y^{-1}]$, the Laurent polynomial ring over $x,y$. From now on, we extend $T^2$ to its complexification $(\mathbb{C}^*)^2$. We extend the function $P(x,y)$ to $(\mathbb{C}^*)^2$ by the Laurent polynomials $p_{ij}(x,y)$ described above.

We still have $P^2=P$ since it is an algebraic relation and is valid on the real torus. The rank of $P$ on the entire $(\mathbb{C}^*)^2$ is always $q$ since $P^2=P$ implies that $\text{rank}(P)=\tr(P)$ and $\tr(P)$ is continuous.

Denote
\begin{equation}
S=\{u\in R^N|Pu=u\},
\end{equation}
which is the $R$-module of global Laurent sections (each component is a Laurent polynomial of $x,y$). 

\emph{Remark:} For any $(x,y)$, $\{u\in \mathbb{C}^N|P(x,y)u=u\}$ is the fiber of $P^*(\tau)$ at $(x,y)$, which is of dimension $q$. The purpose to construct $S$ is that we want to find global basis of the bundle $P^*(\tau)$ made of Laurent polynomials. The below lemma tells us that we can do it at least locally.
 
\emph{Lemma (local structure):} For $\forall (x_0,y_0)\in (\mathbb{C}^*)^2$, there exists $u_1,\cdots,u_q\in S$ such that $u_1(x,y), \cdots, u_q(x,y)$ are linear independent in a neighbourhood of $(x_0,y_0)$.

\emph{Proof of Lemma:} For a fixed $(x_0,y_0)$, choose a basis of $\mathbb{C}^N$ so that $P(x_0,y_0)=\text{diag}(1,\cdots,1,0,\cdots,0)$. Under this basis, we write $1-P(x,y)$ as a block matrix:
\begin{equation}
1-P(x,y)=\frac{1}{x^ay^b}\begin{bmatrix}
    A(x,y)  & B(x,y) \\
    C(x,y)  & D(x,y)
\end{bmatrix},
\end{equation}
where $A,B,C,D$ are matrixes with polynomial elements. By continuity, $D$ is nonsigular on a neighbourhood of $(x_0,y_0)$. On this neighbourhood,
\begin{equation}
\begin{bmatrix}
A& B \\
C  & D
\end{bmatrix}=\begin{bmatrix}
I_q  & B \\
0  & D
\end{bmatrix}
\begin{bmatrix}
A-BD^{-1}C  & 0 \\
D^{-1}C  & I_{N-q}
\end{bmatrix},
\end{equation}
where $I_q$ denotes the identity matrix of size $q$, etc. Since $\text{rank}(1-P)=N-q$, we know $A-BD^{-1}C=0$.  Therefore, a basis of in $\ker(1-P)$ (dim=$q$) is given by the columns of the following $N\times q$ matrix:
\begin{equation}
\begin{bmatrix}
    (\det D)I_q \\
    - (\det D)D^{-1}C
\end{bmatrix}.
\end{equation}
Here we keep $\det(D)$ so that each element of the above matrix is a polynomial (no denominator), as required by the lemma.

\newcommand{\mm}{\mathfrak{m}}
The above lemma and its proof tell us $S$ is a locally free module. Indeed, $(\mathbb{C}^*)^2$ is an affine variety with coordinate ring $R$, so according to the Hilbert's nullstellensatz, each maximal ideal of $R$ corresponds to a point in $(\mathbb{C}^*)^2$ (The correspondence of ideal and point is the basic idea of algebraic geometry. Readers who are not familar with these notions may refer to \cite{reid_2001}.) For the maximal ideal $\mm$ corresponds point $(x_0,y_0)$, consider $S_\mm=\{\frac{s}{f}|s\in S,g\in R,g(x_0,y_0)\neq 0\}$, the localization \cite{reid_2001} of $S$ at $\mm$. Then every $v\in S_\mm$ can be uniquely written as a linear combination of $v_i$ (the image of $u_i$ in $S_\mm$) with coefficients in $R_\mm$. So $S_\mm$ is a free $R_\mm$-module with rank $q$.

Back to the original question. Since $S\subseteq R^N$ and $R$ (as a quotient of a polynomial ring) is a Noetherian ring, $S$ is a Noetherian $R$-module.  Thus $S$ is a projective module\footnote{A finite-generated module over a Noetherian ring is projective iff it is locally free. Moreover, it is enough to verify this for the localization at every maximal ideal. See, for example, \cite{Clark}.}. According to a generalization of the Quillen-Suslin theorem\footnote {A finite-generated projective module over the polynomial ring $k[x_1,\cdots,x_n]$ is free. See, for example, \cite{Lang}} on the Laurent polynomial ring \cite{Swan}, $S$ must be a free module. Fix a basis of $S$, then each element $s_i$ of the basis must be a everywhere-nonzero section, otherwise the lemma breaks down at points where $s_i$ vanishes (as a vector in $\mathbb{C}^N$). So we have found the desired set of global sections. \hfill $\Box$

Geometrically, any such $P_{\vx,\vy}$ give rise to bundle $P^*(\tau)$ with algebraic structure. The Quillen-Suslin theorem confirms that such ``algebraic bundle" over certain base manifold (also with algebraic structure) must be trivial.

Now we can use a continuity argument to prove that as long as $c\neq 0$, $L(P)$ will have a lower bound $\epsilon$ (which may depend on $c$ and $N$). The idea is: if not, there will be a sequence $\{P^i\}$ of maps such that $L(P^i)\to 0$. A limit $\tilde{P}$ of $\{P^i\}$ will satisfy $L(\tilde{P})=0$. However the Chern number should not change and thus nonzero. This contradicts Proposition 1.

\textbf{Proposition 2.} Fix exponent $s>2$, region $F$, Chern number $c\neq 0$, number of orbitals per site $D$ and number of sites per cell $|C|$, then $\exists \epsilon>0$ such that $L(P)>\epsilon$ for $\forall P\in C^{\omega}(T^2,\Gr(N,q))$ ($C^{\omega}$ means real analytic) with Chern number $c(P)=c$.

\textbf{Proof:} If not, there exist a sequence of maps $P^i(\vk)$ with the same Chern number $c^i=c\neq 0$ such that $L(P^i)\to 0$. We Fourier expand each $P^i(\vk)$:
\newcommand{\bvx}{{\bar{\vx}}}
\newcommand{\bvy}{{\bar{\vy}}}
\newcommand{\bF}{\bar{F}}
\begin{equation}
P_{\lambda,\mu}^i(\vk)=\sum_{\bvx\in \bF}P^i_{\bvx,\lambda,\mu}e^{i\vk\cdot \bvx}+\sum_{\bvx\notin \bF}P^i_{\bvx,\lambda,\mu}e^{i\vk\cdot \bvx}.
\end{equation}
Here $\bF$ is a set of cells such that $\vx\notin \cup\bF$ implies $\vx-\vy\notin F$ for $\forall \vy\in C_0$ ($C_0$ is the central unit cell containing \text{0}), see Fig.~\ref{pic-region}, obviously $F\subset \cup \bF$; $\bvx$ means the cell containing $\vx$. One can understand $e^{i\vk\cdot\bvx}$ (and $|\bvx|^s$ in the following) as $e^{i\vk\cdot\vx}$ (and $|\vx|^s$) where $\vx\in\bvx$. 

\begin{figure}
  \centering
  \includegraphics[width=0.4\textwidth]{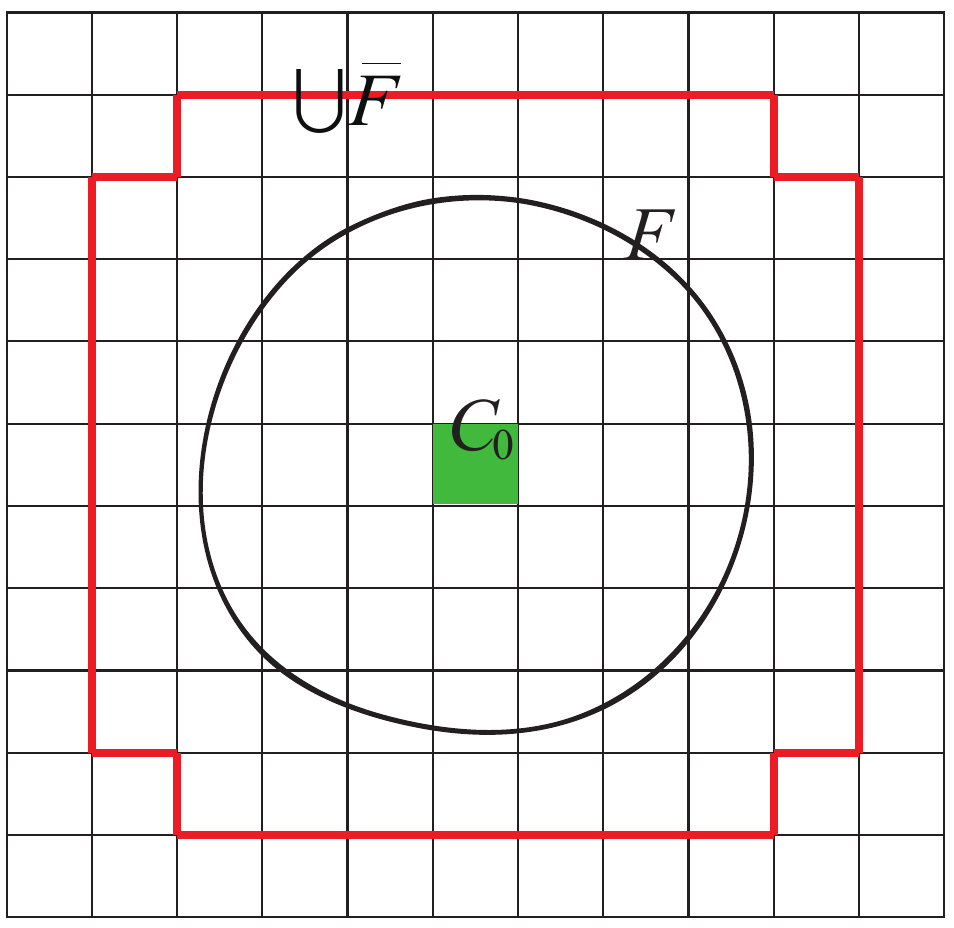}\\
  \caption{The definition of $\bF$. In this figure, the each block is a unit cell, the green one is $C_\textbf{0}$. $F$ is bounded by the black circle and $\cup\bF$ is bounded by the red lines. Roughly speaking, $\cup\bF$ is the extension of $F$ by two cells.}\label{pic-region}
\end{figure}

Since $P_{\lambda,\mu}^i(\vk)$ is uniformly bounded (the Grassmanian is compact), $P^i_{\vx,\lambda,\mu}$ are bounded, thus there is a converging subsequence of $P^i_{\bar{\textbf{0}},\lambda=0,\mu=0}$.
We pick up this subsequence and do the same thing for each point in $\cup\bF$. As a result, we can assume without loss of generality that $P^i_{\bvx,\lambda,\mu}$ converge for $\forall \bvx\in \bF$ and $\forall \lambda,\mu$. Denote the limit as $\tilde{P}_{\bvx,\lambda,\mu}$. Define $\tilde{P}(\vk)$ as
\[\tilde{P}_{\lambda,\mu}(\vk)=\sum_{\bvx\in \bF}\tilde{P}_{\bvx,\lambda,\mu}e^{i\vk\cdot\bvx}.\]

We claim $P^i\rightrightarrows \tilde{P}$ as maps into $(M(N,\mathbb{C}),||\cdot||_{HS})$ ($\rightrightarrows$ means uniformly converge). Indeed, denote $G^i=P^i-\tilde{P}$, using the Cauchy inequality, we have
\begin{equation}\label{eq-Cauchy}
\begin{aligned}
&||G^i(\vk)||^2=||\sum_{\bvx}G^i_\bvx e^{i\vk\cdot \bvx}||^2\leq (\sum_\bvx||G^i_{\bvx}||)^2\\
\leq&(\sum_{\bvx\in \bF}1+\sum_{\bvx\notin \bF} \frac{1}{|\bvx|^{s}})(\sum_{\bvx\in \bF}||G^i_{\bvx}||^2+\sum_{\bvx\notin \bF}|\bvx|^{s}||G^i_{\bvx}||^2).
\end{aligned}
\end{equation}

The first factor converges when $s>2$. The second factor converges to 0 since $F$ is finite, $G^i_{\bvx}=P^i_{\bvx}-\tilde{P}_\bvx\to 0$ for $\forall \bvx\in \bF$ by construction, and 
\begin{equation}
\begin{aligned}
&\sum_{\bvx\notin \bF}|\bvx|^{s}||G^i_{\bvx}||^2=\sum_{\bvx\notin \bF,\lambda,\mu}|\bvx|^{s}||G^i_{\bvx,\lambda,\mu}||^2\\
= &\sum_{\vy\in C,\vx\notin \cup\bF}|\bvx|^{s}||G^i_{\vx,\vy}||^2 \lesssim \sum_{\vy\in C,\vx\notin \cup\bF}|\vx-\vy|^{s}||G^i_{\vx,\vy}||^2\\
\leq & |C|L(P^i)\to 0.
\end{aligned}
\end{equation}
(``$\lesssim$" means less than the right hand side times a constant which only depends on $s$ and $F$. Since $\vx\notin \cup\bF$ implies $\vx-\vy\notin F$, such constant exists). So the right hand side of Eq. (\ref{eq-Cauchy}) converges to 0, uniformly with respect to $\vk$.

It is easy to show $\tilde{P}(\vk)\in \Gr(N,q)$ for $\forall \vk$, so $\text{Im}(\tilde{P})\subseteq \Gr(N,q)$ and we can define its Chern number 
$\tilde{c}$. According to Proposition 1, $\tilde{c}=0$. 

On the other hand, we claim that $\tilde{c}=\lim c^i=c$. Indeed, since $P^i\rightrightarrows \tilde{P}$, there $\exists I$ such that $|P^i(\vk)-\tilde{P}(\vk)|<\delta$ when $i>I$ ($\delta$ is chosen small enough as follows). Since $\Gr(N,q)$ is compact, there exists a small $r$ such that if we pick up the disk with radius $r$ in the normal subspace of each point $p\in \Gr(N,q)$, the normal disks do not intersect each other (a ``tubular neighborhood" of $\Gr(N,q)$ in $M(N,\mathbb{C}$). Also due to compactness, we can choose $\delta$ small enough so that whenever $p,p'\in \Gr(N,q)$ and $|p-p'|<\delta$, each point in the segment between $p,q$ belongs (and only belongs) to a unique normal disk. Now we can construct a homotopy $\Psi(t,\vk)$ between $P^i$ and $\tilde{P}$ as: $\Psi(t,\vk)=\text{the center of the disk that~} (1-t)P^i(\vk)+t\tilde{P}(\vk) \text{~belongs to}$. Here, the $``+"$ is the addition of matrices.
Thus, $\tilde{c}=c^i$ when $i>I$, which means $\tilde{c}=c\neq 0$. 

This contradiction shows the existence of the lower bound. \hfill $\Box$

\bibliography{merabound}